\documentclass[12pt]{article}
\pdfoutput=1
\usepackage{amsmath,amssymb,amsthm,bm,graphicx,float,array,multirow,multicol,rotfloat,caption,subcaption,hyperref,cleveref,enumerate,geometry,mathdots,adjustbox,booktabs,parskip,mathtools,tikz,tikz-cd,pdflscape,csquotes,lscape,rotating,empheq,mathrsfs}
\usepackage[para]{threeparttable}
\usepackage[all]{xy}
\usepackage[normalem]{ulem}
\usepackage[numbers,sort&compress]{natbib}
\usepackage{physics}
\usepackage{algorithm2e}
\RestyleAlgo{ruled}

\usetikzlibrary{positioning}
\numberwithin{equation}{section}
\numberwithin{figure}{section}
\theoremstyle{remark}
\allowdisplaybreaks
\makeatletter

\usetikzlibrary{arrows,shapes.misc,positioning,decorations.pathmorphing,decorations.markings,decorations.pathreplacing,matrix,patterns,backgrounds}
\tikzset{rectangle/.style={draw,circle,inner sep=1pt}, 
    big arrow/.style={decoration={markings,mark=at position 1 with {\arrow[scale=1.5,#1]{>}}}, postaction={decorate}, shorten >=0.4pt}, 
    scale cd/.style={every label/.append style={scale=#1}, cells={nodes={scale=#1}}},
    brace/.style={decoration={brace, mirror},decorate}}

\tikzset{
node/.style={circle, thick, draw=black!100,fill=white!100,  minimum size=6mm, inner sep=0pt},
sonode/.style={circle, thick, draw=black!100,fill=red!100,  minimum size=3mm, inner sep=0pt},
spnode/.style={circle, thick, draw=black!100,fill=blue!100,  minimum size=3mm, inner sep=0pt},
fnode/.style={rectangle, thick, draw=black!100,fill=white!100,  minimum size=3mm, inner sep=0pt},
tnode/.style={rounded rectangle, outer sep=0pt, thick, minimum size=5mm}
}

\theoremstyle{plain}
\newtheorem*{thm*}{Theorem}

\theoremstyle{definition}

\newtheorem*{defn*}{Definition}

\makeatother

\begin{document}

\begin{titlepage}
\vspace*{-3cm} 
\begin{flushright}
{\tt CALT-TH-2023-048}\\
\end{flushright}
\begin{center}
\vspace{2.2cm}
{\LARGE\bfseries Central extensions of higher groups:\\ Green--Schwarz mechanism and 2-connections}
\vspace{1.6cm}

{\large
Monica Jinwoo Kang$^{1\,2}$ and Sungkyung Kang$^{3}$\\}
\vspace{.7cm}
{ $^1$ Department of Physics and Astronomy, University of Pennsylvania\\
Philadelphia, PA 19104, U.S.A.}\par
\vspace{.2cm}
{ $^2$ Walter Burke Institute for Theoretical Physics, California Institute of Technology}\par
{Pasadena, CA 91125, U.S.A.}\par
\vspace{.2cm}
{ $^3$ Mathematical Institute, University of Oxford}\par
{Oxford, OX2 6GG, United Kingdom}\par
\vspace{.6cm}

{\tt monica6@sas.upenn.edu, kangs@maths.ox.ac.uk}\par
\vspace{1.6cm}
\textbf{Abstract}
\end{center}

\noindent We study the smooth $2$-group structure arising in the presence of quantum field theory with one-form symmetry. We acquire $2$-group structures obtained by a central extension of the zero-form symmetry by the one-form symmetry. We determine that the existence of a $2$-group structure is guaranteed by Chern--Simons levels. We further verify how we will be able to provide a fix to the current $2$-group problems by using the bibundle model. We outline the principal $2$-connection theory with respect to such $2$-group and compare it with the ansatz obtained from the Green--Schwarz mechanism. We further propose the existence of smooth $\infty$-group symmetries in quantum field theory.

\vfill 
\end{titlepage} 

\tableofcontents

\newpage

\section{Introduction}

Symmetry has been one of the most central feature that provides much toolkit to study the structure of quantum field theories. Quantum field theory can admit one-form symmetries and higher-form symmetries, sourced by their corresponding currents. It is suggested in \cite{sharpe2015notes} that these symmetries can form mathematical objects as special cases of $2$-group or higher-group symmetries. 
While some 2-group symmetries found in physics are discrete, most of them have topological, or even smooth, aspects, which predicts a rich higher differential-geometric structures underlying those symmetries. Among them, Chern--Simons theory is probably one of the most naturally occurring and interesting sources of smooth 2-groups \cite{sharpe2015notes}, as \cite{cordova2019exploring} claim the presence of 2-groups in 3d Chern--Simons theory. Given any compact Lie group $G$, the level of a 3d Chern--Simons theory with gauge group $G$ defines via Chern-Weil homomorphism a cohomology class in $H^4(BG;\mathbb{Z})$, which then induces a uniquely defined 2-group which is formed by a central extension of the form
\begin{align}
    0\rightarrow BU(1)\rightarrow \tilde{G}\rightarrow G\rightarrow 1.
\end{align}

Surprisingly, such 2-groups were already of independent interest in mathematical literature. When $G$ is the spin group $\mathrm{Spin}(n)$, then the extension $\tilde{G}$ is called the \emph{string group} $\mathrm{String}(n)$, which were originally constructed as the 3-connected cover of $\mathrm{Spin}(n)$, i.e. the term that arises next to $\mathrm{Spin}(n)$ in the Whitehead tower of $O(n)$. Since $\pi_3(\tilde{G})$ is nonzero, it is clear that $\tilde{G}$ cannot be represented as a finite-dimensional Lie group, and hence it has drawn much interest for decades from the purely mathematical viewpoint; see \cite{stolz1996conjecture,stolz2004elliptic} for some early approaches on constructing geometric models of $\mathrm{String}(n)$. From this, the framework of smooth 2-groups and smooth $\infty$-groups, as well as the notion of principal bundles and connections with respect to those ``higher'' gauge groups (see \cite{demessie2017higher,nikolaus2015principal,nikolaus2015principal2} for more details in this direction), had been developed.

Starting from \cite{sharpe2015notes}, there has recently been a surge of interest on 2-group symmetries in physics literature. In most of them, 2-group symmetries are found by applying Green--Schwarz anomaly cancellation mechanism to 0-form and 1-form symmetries of quantum field theories. This approach then yields a transformation rule for $A$ and $B$ fields in which 0-form and 1-form symmetries are mixed together. For example, Tachikawa derived in \cite{tachikawa2020gauging} a discrete 2-group symmetry by taking a quotient of a field theory which has a nontrivial anomaly by a subgroup of its symmetry group. Cordova--Dumitrescu--Intriligator follows a different approach involving Chern--Simons theory in \cite{cordova2019exploring} by mixing a 0-form $SU(n)$-symmetry and a 1-form $U(1)$-symmetry using a nontrivial Chern--Simons level $\kappa\in \mathbb{Z}$.

While both approaches have their own benefits, there are some downsides to them. For Tachikawa's approach, it is clear that it can only be applied to finite group symmetries, and thus the resulting 2-group is also ``finite''; while its derivation seems to be more formal and rigorous, its finiteness gives it a fundamental limitations on its possible applications. For example, having a finite 2-group gauge theory on a spacetime is equivalent to having a 2-representation of its fundamental 2-group, which implies that the theory is inherently topological and thus cannot produce any invariant which makes use of smooth structures. In particular, it will be unable to produce something similar to Seiberg-Witten invariants, which are the most powerful tools for distinguishing smooth structures on 4-manifolds which are homeomorphic, but not diffeomorphic, to each other. 

Things are quite different for Cordova--Dumitrescu--Intriligator: while it arises naturally from Chern--Simons theory and leads to potentially interesting 2-gauge theory using the corresponding smooth 2-group, his arguments are less rigorous. While they claim to construct a smooth 2-group (note that smooth structure is necessary here in order for a gauge theory to exist) induced by central extensions of the form
\[
1 \rightarrow BU(1) \rightarrow \tilde{G} \rightarrow G \rightarrow 1
\]
corresponding to a cohomology class 
\begin{align}
    \alpha \in H^3(G,U(1)),
\end{align}
one should be very careful about this cohomology; if one uses smooth chains of the form 
\begin{align}
    G^n\rightarrow U(1), 
\end{align}
then the cohomology is zero, as already pointed out in \cite{baez2004higher}. This needs a further fix which was first figured out by Schommer--Pries in \cite{schommer2011central}.

One further issue to note here is that the transformation rules one gets via Green--Schwarz mechanism in the case of Cordova--Dumitrescu--Intriligator does not quite match the one predicted from a mathematically rigorous foundation of 2-connections on principal smooth 2-group bundles; in fact, they match only in very special cases. Thus, in the cases of both Tachikawa and Cordova--Dumitrescu-Intriligator, it seems that going beyond just the existence of interesting 2-groups, and developing new physics which cannot be simply derived from 0-form and 1-form symmetry groups, is currently a difficult task. 

Fortunately, the non-rigorousness of the existence of 2-groups considered by Cordova--Dumitrescu--Intriligator admits a mathematically rigorous fix, which has been known in the mathematics literature for a decade; we will  outline this ``fix'' in this paper, but we give a very short but intuitive overview here. Rather than using the model of semistrict weak Lie 2-group of Baez--Lauda, we have to use a more generalized model, given by the bibundle model of ``smooth 2-groups'' by Schommer--Pries. Then it follows that, given a Lie group $G$ and an abelian Lie group $A$, central extensions of $G$ by $BA$ are classified by elements not by ``naive group cohomology'' given by considering smooth group cochains, but rather in $H^3_{diff}(G,A)$, where $H^3_{diff}$ denotes the Segal--Mitchinson--Brylinski differential group cohomology. When $G$ is compact and $A$ is $U(1)$, it is known that $H^3_{diff}(G,A)$ is isomorphic to the singular cohomology $H^4(BG,\mathbb{Z})$, whose elements correspond to levels in 3d Chern--Simons theory. This condition covers the cases considered in \cite{cordova2019exploring}.

Then, by tracking the proof of Schommer--Pries' classification theorem, it is possible to explicitly describe the bibundle model describing the central extension 2-groups in question, which we can then convert to the (non-semistrict) weak Lie 2-group model and Lie-differentiate to describe its Lie 2-algebra. The resulting Lie 2-algebra admits an extremely simple model as a 2-term $L_\infty$-algebra whose terms consist of $\mathfrak{u}(1)$ and the Lie algebra of $G$. Then we use this simplified Lie 2-algebra to explicitly describe the theory of principal 2-connections for smooth 2-groups of Cordova--Dumitrescu--Intriligator and compare it with infinitesimal gauge 2-group transformation rules predicted from applying Green--Schwarz mechanism. It follows that they match only in very special cases, and the matching condition is equivalent to the 2-group gauge theory admitting ``2-monodromies'' along surfaces.

Given this, a natural question arise: can we realize smooth (non-strict) 2-group, and in general, $\infty$-group symmetries in physics? Tachikawa's approachin \cite{tachikawa2020gauging}, albeit being in a discrete realm, goes directly in this direction, and it seems that the same should also be possible for higher smooth groups such as $\mathrm{String}(n)$ or 2-groups of Cordova--Dumitrescu--Intriligator, but currently it is very unclear. It can also be asked whether 2-groups, smooth or discrete, can be realized as global symmetry groups acting on spacetimes, or some resolved versions of them.

Furthermore, we can also ask whether smooth $\infty$-groups can also arise in physics in an approach similar to the ones taken by Cordova--Dumitrescu--Intriligator. Since they rely on the basic idea of applying Green--Schwarz mechanism, which is the same as the idea used by Tachikawa, and Tachikawa's result imply that discrete $n$-groups can arise as symmetries of field theories for any $n\ge 1$, it seems natural to assume that smooth $n$-group symmetries should also exist. Furthermore, we can go a step further and ask whether smooth $\infty$-group symmetries can also exist in nature.


This paper is organized as follows. In section 2, we recall two models for 2-groups with smooth structures, which are weak Lie 2-groups and smooth 2-groups. In section 3, we recall the definition of central extensions of smooth 2-groups and explain Schommer--Pries' result on classifying such extensions in terms of elements in $H^3_{diff}$. Finally, in section 4, we explicitly describe infinitesimal gauge transformation rules for 2-groups obtained by nontrivial central extensions of $U(1)$ and $SU(n)$ by $BU(1)$ and compare them with the ones given in \cite{cordova2019exploring}.


\section{Smooth 2-groups}

In this section, we review the definitions of 2-groups (with smooth structures) due to Baez--Lauda (which defines \emph{Lie 2-groups}) and Schommer--Pries (which defines \emph{smooth 2-groups}), which are modelled using Lie groupoids and bibundles, respectively. We will closely follow the works of \cite{baez2004higher} and \cite{schommer2011central}.

In short, Lie 2-groups are 2-group objects internal to the strict 2-category $\mathbf{LieGpd}$ of Lie groupoids, smooth functors (i.e. maps between Lie groupoids) and natural transformations. On the other hand, smooth 2-groups are 2-group objects internal to the weak 2-category $\mathbf{Bibun}$ of bibundles, which is an enhancement of $\mathbf{LieGpd}$. We will describe here the definitions of 2-group objects and the weak 2-category $\mathbf{Bibun}$; this will be sufficient for readers to figure out the definitions of Lie 2-groups and smooth 2-groups.

\subsection{2-group objects internal to a weak 2-category}
Given a weak 2-category $\mathcal{C}$ with finite products, a \emph{2-group object in $\mathcal{C}$} is a following collection of data:
\begin{itemize}
    \item An object $C_0$ of $\mathcal{C}$;
    \item 1-morphisms $m:C_0\times C_0\rightarrow C_0$ and $e:\ast\rightarrow C_0$, where $\ast$ denotes the terminal object of $\mathcal{C}$ (which exists because $\mathcal{C}$ has finite products);
    \item Invertible 2-morphisms $a:m\circ (m\times 1)\rightarrow m\circ (1\times m)$, $\ell:m\circ (e\times 1)\rightarrow \simeq$, and $r:m\circ (1\times e)\rightarrow \simeq$, where $\simeq$ denotes the isomorphisms $\ast\times C_0\simeq C_0\simeq C_0\times\ast$.
\end{itemize}
Here, the following conditions should be satisfied.
\begin{itemize}
    \item $(pr_1,m):C_0\times C_0 \rightarrow C_0\times C_0$ is an equivalence, where $pr_1$ denotes the projection morphism onto the first copy of $C_0$;
    \item $a,\ell,r$ satisfy the pentagon and triangle identities drawn below.
\end{itemize}
\[
\xymatrixrowsep{.3in}
\xymatrixcolsep{.1in}
\xymatrix{
&& (m\times m)\circ m\ar[rrdd]^{(1\times 1\times m)\circ a}\\
\\
(m\times 1\times 1)\circ (m\times 1)\circ m\ar[uurr]^{(m\times 1\times 1)\circ a} \ar[dd]_{(a\times 1)\circ m}&&&& (1\times 1\times m)\circ (1\times m)\circ m \\
\\
 (1\times m\times 1)\circ (m\times 1)\circ m\ar[rrrr]^{(1\times m\times 1)\circ a} &&&& (1\times m\times 1)\circ (1\times m)\circ m\ar[uu]_{(1\times a)\circ m}
}
\]

\[
\xymatrix{
(1\times e\times 1)\circ (m\times 1)\circ m \ar[rr]^{(1\times e\otimes 1)\circ a}\ar[dr]_{(r\times 1)\circ m} & & (1\times e\times 1)\circ (1\times m)\circ m \ar[ld]^{(1\times \ell)\circ m} \\
& m
}
\]

This definition, while seemingly daunting, admits the following intuitive interpretation. Recall that a group, i.e. 1-group, is a set with an associative multiplication structure where every element is invertible. This can be thought of as a category with one object. A group is given a structure of a Lie group if the multiplication structure and the inversion map are smooth.

The intuitive idea of 2-groups is given by ``relaxing'' the group laws, so that the group laws hold only up to a certain notion of ``equivalences''. The 1-morphisms $m$ and $e$ can be interpreted as the ``multiplication structure'' and the ``unit''. Note here that, under this interpretation, we still have that multiplications of two group elements are uniquely determined. This is a nice feature of Lie 2-groups, but is also a weakness of it. In general, in the definition of smooth 2-groups via Kan simplicial manifolds (which is equivalent to the bibundle model of Schommer--Pries), which is probably the most general definition of 2-groups with smooth structures, multiplication (i.e. composition of 1-morphisms) need not be uniquely determined; it is unique only up to 2-morphisms.

Continuing our discussion of 2-group structures, the 2-morphism $a$ is the ``associator''; this is the central feature of 2-groups. It arises because the associativity law now holds only up to equivalence, and the equivalence here is specified by $a$. The 2-morphisms $\ell$ and $r$ also have similar interpretations; they come from requiring that the identity, i.e. $e$, is no longer the strict multiplicative identity, but rather an identity up to equivalence. Multiplication on the left and right give $\ell$ and $r$, respectively.

Given this interpretation, the pentagon and triangle identities now follow naturally. The triangle identity follows from the ambiguity of evaluating the triple product $g_1\cdot 1\cdot g_2$ to $g_1g_2$ as follows.
\[
\xymatrix{
(g_11)g_2 \ar[rr]\ar[dr] & & g_1(1g_2) \ar[ld] \\
& g_1g_2
}
\]
Furthermore, the pentagon identity follows from the ambiguity of evaluating the quadruple product $g_1 g_2 g_3 g_4$, as follows. In other words, the pentagon here is the 2-dimensional associahedron $K_4$.
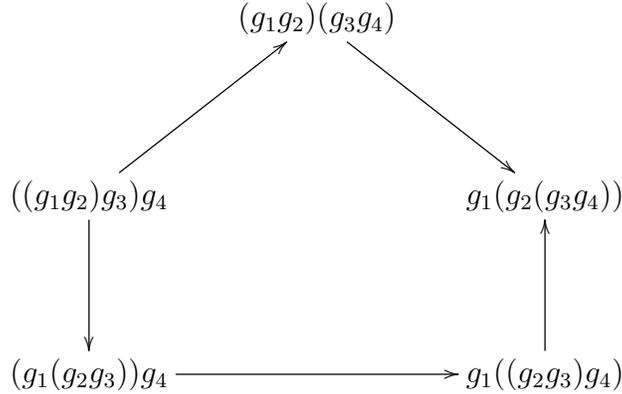
\[
\xymatrixrowsep{.3in}
\xymatrixcolsep{.1in}
\xymatrix{
&& (g_1g_2)(g_3g_4)\ar[rrdd]\\
\\
((g_1g_2)g_3)g_4\ar[dd]\ar[rruu]&&&& g_1(g_2(g_3g_4)) \\
\\
(g_1(g_2g_3))g_4 \ar[rrrr]&&&& g_1((g_2g_3)g_4)\ar[uu]
}
\]

\subsection{The weak 2-category $\mathbf{Bibun}$}
Recall that a \emph{Lie groupoid} is a collection of following data (satisfying certain conditions which we do not describe here):
\begin{itemize}
    \item Smooth manifolds $G_0$ and $G_1$;
    \item Surjective submersions $s,t:G_1\rightarrow G_0$ (seen as \emph{source} and \emph{target} maps);
    \item A smooth map $\circ:G_1\times G_1\rightarrow G_1$ (seen as \emph{composition});
    \item A smooth map $e:G_0\rightarrow G_1$ (seen as \emph{identity}).
\end{itemize}

The notion of Lie 2-groups, whose definition is precisely the one discussed in the previous subsections where ``objects'' are given by Lie groupoids, ``1-morphisms'' are given by functors between Lie groupoids, and ``2-morphisms'' are given by natural transformations between them, has the advantage of being very simple and explicit. However its property that any composable pairs of 1-morphisms have unique compositions makes the definition not strong enough to deal with sophisticated examples, like the central extensions of Lie groups by $BU(1)$ which we will discuss in the next section. 

There has been several attempts to remedy this problem inside the framework of Lie 2-groups. one notable attempt was the construction of a model of string 2-groups \cite{baez2007loop}, i.e. certain central extensions of spin groups by $BU(1)$, as Lie 2-groups where we use infinite-dimensional Frech\'{e}t manifolds (involving extensions of loop groups) instead of finite-dimensional manifolds. Henriques' integration of $L_\infty$-algebras \cite{henriques2008integrating} also uses a similar set of ideas.

The first finite-dimensional model of string 2-groups was given in by Schommer--Pries in \cite{schommer2011central}. He uses the notion of bibundles to construct a weak 2-category whose objects are Lie groupoids but has more 1-morphisms than just functors between Lie groupoids. This model, which we call the \emph{bibundle model}, gives a more flexible and general definition of ``2-groups with finite-dimensional smooth structures'', which he calls as \emph{smooth 2-groups}. This definition allowed him to prove the classification theorem for central extensions of Lie groups by $BU(1)$ via some version of smooth group cohomology theory, which will be briefly reviewed in the next section.

In this section, we will briefly recall the definition of the bibundle 2-category, $\mathbf{Bibun}$; combining this with the definitions in the previous subsection gives the complete definition of smooth 2-groups via bibundle model. In order to present the definition of the weak 2-category $\mathbf{Bibun}$, we have to define bibundles. A \emph{principal bibundle} from a Lie groupoid $H=(H_1\rightrightarrows H_0)$ to a Lie groupoid $G=(G_1\rightrightarrows G_0)$ is a collection of following data:
\begin{itemize}
    \item A smooth manifold $B$;
    \item A smooth map $\tau:B\rightarrow G_0$;
    \item A surjective submersion $\sigma:B\rightarrow H_0$;
    \item Smooth maps $G_1\times^{s,\tau}_{G_0}B\rightarrow B$ and $B\times^{\sigma,t}_{H_0}B\rightarrow H_1$, called \emph{left/right action maps} and denoted like group actions.
\end{itemize}
The above data is subject to the following conditions.
\begin{itemize}
    \item $g_1(g_2b)=(g_1g_2)b$ for $(g_1,g_2,b)\in G_1\times^{s,t}_{G_0}G_1\times^{s,\tau}_{G_0}B$;
    \item $(bh_1)h_2=b(h_1h_2)$ for $(b,h_1,h_2)\in B\times^{\sigma,t}_{H_0}H_1\times^{s,t}_{H_0}H_1$;
    \item $b\mathrm{id}_H(\sigma(b))=b$ and $\mathrm{id}_G(\tau(b))b=b$ for $b\in B$;
    \item $g(bh)=(gb)h$ for $(g,b,h)\in G_1\times^{s,\tau}_{G_0}B\times^{\sigma,t}_{H_0}H_1$;
    \item The map $(g,b)\mapsto (gb,b):G_1\times^{s,\tau}_{G_0}B\times B\times^{\sigma,\sigma}_{H_0}B$ is an diffeomorphism.
\end{itemize}

\begin{align}
\xymatrix{
G_1\ar@<-.5ex>[dd]_{s}\ar@<.5ex>[dd]^{t} && && H_1\ar@<-.5ex>[dd]_{s}\ar@<.5ex>[dd]^{t}\\
&& B\ar[lld]_{\tau}\ar@{->>}[rrd]^{\sigma} \\
G_0 && && H_0
}
\end{align}

A bibundle map between (principal) bibundles $B$ and $B^\prime$ between Lie groupoids $H$ and $G$ is a smooth map $B\rightarrow B^\prime$ which is invariant under structure maps (i.e. $\sigma$ and $\tau$) of $B$ and $B^\prime$.

Now the weak 2-category $\mathbf{Bibun}$ is defined as follows.
\begin{itemize}
    \item Objects of $\mathbf{Bibun}$ are Lie groupoids;
    \item 1-morphisms of $\mathbf{Bibun}$ are bibundles between Lie groupoids;
    \item 2-morphisms of $\mathbf{Bibun}$ are bibundle maps.
\end{itemize}
It is worth noting here that any smooth map between Lie groupoids can be regarded as a bibundle by a procedure called \emph{bundlization}, as follows. Given a functor $\phi=(\phi_0,\phi_1)$ between Lie groupoids $H$ and $G$, i.e. $\phi_i:H_i\rightarrow G_i$ for $i=0,1$, the bundlization of $\phi$ is given by the bibundle
\[
B = H_0 \times_{G_0}^{\phi_0,s} G_1,
\]
together with the structure maps
\[
\begin{split}
\tau &: H_0 \times_{G_0}^{\phi_0,s} G_1 \xrightarrow{proj} G_1 \xrightarrow{t}G_0, \\
\sigma &: H_0 \times_{G_0}^{\phi_0,s} G_1 \xrightarrow{proj} H_0.
\end{split}
\]
Furthermore, the action maps are given as
\[
g^\prime(x,g)=(x,g^\prime \circ g),\quad (x,g)h = (s(h),g\circ \phi_1(h))
\]
for $g,g^\prime\in G_1$, $h\in H_1$, and $(x,g)\in B$. This gives a way to interpret functors between Lie groupoids as bibundles, and hence the strict 2-category $\mathbf{LieGpd}$ is a subcategory of $\mathbf{Bibun}$. In other words, the concept of smooth 2-groups is an enhancement of the concept of Lie 2-groups.

\subsection{Strict Lie 2-groups, i.e. crossed modules}
\label{sec:StrictLie2Group}

A very special case of a Lie 2-group is when it is \emph{strict}, which is equivalent to saying that it is induced by a \emph{crossed module} of Lie groups. Since such Lie 2-groups will arise later, we briefly recall the relevant definitions here.

We start with the definition of crossed modules. A (smooth) crossed module is a quadruple $(G,H,t,\alpha)$ where $G,H$ are Lie groups, $t:H\rightarrow G$ is a smooth homomorphism, and $\alpha:G\rightarrow \mathrm{Aut}(H)$ is a smooth action of $G$ on $H$ by automorphisms. This data should satisfy the \emph{Peiffer identity}
\[
\alpha(t(h),h^\prime)=hh^\prime h^{-1},
\]
i.e. $\alpha\circ t$ is the conjugation action. 

From a smooth crossed module, we can build up a Lie 2-group as follows. We consider the Lie groupoid
\[
G\times H\rightrightarrows G,
\]
with the structure maps $s,t,\circ,e$ given by 
\[
s(g,h)=g,\quad t(g,h)=t(g)h,\quad (t(h)g,h^\prime)\circ (g,h)=(g,h^\prime h),\quad e(g)=(g,1).
\]
Then we define the multiplication map $m$ as
\[
m((g,h),(g^\prime,h^\prime))=(gg^\prime,h(\alpha(g)(h^\prime)))\text{ on } G\times H \text{ and } m(g,g^\prime) = gg^\prime \text{ on }G,
\]
and the unitor $e$ as $e(\ast)=1$. The 2-morphisms $a,\ell,r$ are identity natural transformations; this is why the resulting Lie 2-group is called strict. It is known that the categories of crossed modules and strict Lie 2-groups are equivalent \cite{baez2004higher}, so one can safely say that crossed modules are \emph{precisely} strict Lie 2-groups.

\section{Central extension of Lie groups by $BU(1)$}

The simplest interesting examples of smooth 2-groups which are not strict, i.e. does not come from crossed modules of Lie groups, come from central extensions, as it allows us to construct non-strict Lie 2-groups from strict Lie 2-groups. Since $BU(1)$ is the simplest possible nontrivial strict Lie 2-group which is not equivalent to a Lie group, we are naturally led to consider nontrivial central extensions of the form
\[
0\rightarrow BU(1)\rightarrow H\rightarrow G\rightarrow 1,
\]
where $G$ is a Lie group. When $G$ is a spin group, i.e. $G=\mathrm{Spin}(n)$, then $H$ is called the \emph{string 2-group} $\mathrm{String}(n)$, which is one of the most well-studied smooth 2-groups in mathematical literature, starting probably from \cite{killingback1987world}. Surprisingly, this construction, where $G$ is allowed to be any compact Lie group which arises as a gauge group in various QFTs, recently also gained attention in physics literatures, starting from \cite{sharpe2015notes} and also \cite{cordova2019exploring}. In this section, we will review some definitions and classification results regarding central extensions of Lie groups by $BU(1)$, as we feel that it is now necessary to clarify these.

\subsection{A note on the strict Lie 2-group $BU(1)$}

Before we proceed to central extensions, we start with a brief discussion of the structure of $BU(1)$. Recall from Section \ref{sec:StrictLie2Group} that a (smooth) crossed module is a quadruple $(G,H,t,\alpha)$ where $G,H$ are Lie groups, $t:H\rightarrow G$ is a smooth homomorphism, and $\alpha:G\rightarrow \mathrm{Aut}(H)$ is a smooth action of $G$ on $H$ by automorphisms. Given a crossed module $(G,H,t,\alpha)$, one can associate to it a strict Lie 2-group modelled on the Lie groupoid $G\times H\rightrightarrows G$.

Given any Lie group $G$, one can form a crossed module, denoted $BG$, as follows.
\begin{itemize}
    \item $t$ is given by $t:G\rightarrow 1$;
    \item $\alpha$ is given by the trivial action of $1$ on $G$ by identity.
\end{itemize}
This defines a strict 2-group $BG$, modelled on the Lie groupoid
\[
G\rightrightarrows 1.
\]
When $G$ is abelian, $BG$ is, obviously, also abelian. This allows us to consider central extensions by $BA$ whenever $A$ is an abelian Lie group.

In physical terms, this construction can be interpreted as follows. If we have a $G$-symmetry, it means that we have a 0-form symmetry on the given field theory which forms a group isomorphic to $G$. Now, if we have a $BG$-symmetry, it means that we have a 1-form symmetry on the given theory, instead of 0-form symmetries, forming a group isomorphic to $G$. In more mathematical terms, this means that infinitesimal gauge transformations on a principal $BU(1)$-connection on a manifold $M$ are given by $\mathfrak{u}(1)$-valued 1-forms on $M$. This viewpoint will be discussed more in Section \ref{sec:GS}.

\subsection{Central extensions of discrete groups}

Recall the definition of central group extensions: given a group $G$ and an abelian group $A$, we say that $1\rightarrow A\rightarrow G^\prime \rightarrow G \rightarrow 1$ is a \emph{central extension} of $G$ by $A$ if it is exact and $A$ is contained in the center $Z(G^\prime)$ of $G^\prime$. Two extensions $G^\prime_1$ and $G^\prime_2$ are \emph{equivalent} if there exists a group homomorphism $f:G^\prime_1 \rightarrow G^\prime_2$ which fits into the following commutative diagram.
\[
\xymatrix{
1\ar[r]& A\ar[r]\ar[d]^=& G^\prime_1\ar[r]\ar[d]^f& G\ar[r]\ar[d]^=& 1 \\
1\ar[r]& A\ar[r]& G^\prime_2 \ar[r]& G\ar[r]& 1
}
\]
The equivalence classes of central extensions of $G$ by $A$ are classified by elements of the second group cohomology $H^2 (G,A)$. In particular, to specify a central extension, one only has to specify a cohomology class in $H^2 (G,A)$.

This phenomenon extends naturally to the case of central extensions of \emph{discrete} 2-groups, i.e. when no extra structure is endowed. Given a discrete 2-group $G$, one can extract from it the following invariants:
\begin{itemize}
    \item The group of 1-automorphisms, denoted $\pi_1(G)$;
    \item The group of 2-automorphisms of the identity 1-morphism (which is always abelian), denoted $\pi_2(G)$ (note that $\pi_1(G)$ has a natural action on $\pi_2(G)$);
    \item A homology class $\alpha\in H^3(\pi_1(G),\pi_2(G))$ given by the associator (which becomes a group 3-cocycle). Note that this is the \emph{Sinh invariant} of $G$, defined by Sinh in her doctoral thesis \cite{sinh1975gr}.
\end{itemize}
Conversely, it is known that the data $(\pi_1(G),\pi_2(G),\alpha)$ determines $G$ completely up to equivalence. If $H$ is a discrete group and $A$ is an abelian discrete group, then any central extension (which will be defined in the next section)
\[
0\rightarrow BA\rightarrow G\rightarrow H\rightarrow 1
\]
will satisfy $\pi_1(G)= H$ and $\pi_2(G)=A$, so the associator class $\alpha$ determines $G$. 

This approach was used in several physics literatures to deal with 2-groups, including \cite{cordova2019exploring,tachikawa2020gauging}. Unfortunately, this approach breaks down when the 2-groups that we are dealing with are endowed with extra structures, especially if we want them to be internal to topological spaces or smooth manifolds. This point is indeed very important, and will be discussed in the next subsection.

\subsection{Smooth structures}
Things are, unfortunately, more complicated with extra structures involved; probably the best way to illustrate the difference is to observe what happens for topological groups. Given a topological group $G$ and an abelian topological group $A$, we say that $1\rightarrow A\rightarrow G^\prime \rightarrow G\rightarrow 1$ is a central extension if it is indeed a central extension of groups and all maps involved are continuous and open onto their images. Then, for any nontrivial simple compact Lie group $G$, there exists a nontrivial central extension of the form
\[
1\rightarrow U(1)\rightarrow H \rightarrow LG \rightarrow 1,
\]
and it is well-known that this extension cannot be represented by any element in the second cohomology of continuous group cochains from $LG$ to $U(1)$. In fact, this extension is represented by the ``infinitesimal 2-cocycle''
\[
c:L\mathfrak{g}\times L\mathfrak{g}\rightarrow \mathfrak{u}(1)=i\mathbb{R},\quad c(\alpha,\beta) =  i\int_0^1 \left< \alpha(t),\beta^\prime(t)\right>dt,
\]
where $\left< - , - \right>$ denotes a Killing form on $\mathfrak{g}$, and this does not integrate to a continuous 2-cocycle on $LG\times LG$.

This suggests that the ``naive'' group cohomology defined via continuous group cochains is not the right cohomology theory to use anymore, and this should also be the case in the smooth world. The correct notion of group cohomology for classifying central extensions of topological groups is the Segal-Mitchinson cohomology $H^\ast_{SM}$. This claim can be made precise as follows: given compactly generated topological groups $G$ and $A$ such that
\begin{itemize}
    \item $G$ is paracompact;
    \item $A$ is locally contractible, Hausdorff, and abelian;
\end{itemize}
then equivalence classes of topological central extensions of $G$ by $A$ are classified by elements of $H^2_{SM}(G,A)$ \cite[Section 4]{segal1970cohomology}. 

This cohomology was then extended to the differential group cohomology $H^\ast_{diff}$ by Brylinski \cite{brylinski2000differentiable}. Given Frech\'{e}t Lie groups $A,G,H$, where $A$ is abelian, we say that an exact sequence
\[
1\rightarrow A\rightarrow H\rightarrow G\rightarrow 1
\]
is a central extension if it is a central extension of discrete groups and the map $H\rightarrow G$ is a locally trivial smooth principal $A$-fibration. It is known \cite{brylinski2000differentiable} that such extension classes are classified by the second differential group cohomology, i.e. elements in $H^2_{diff}(G,A)$.

With this phenomenon in mind, we now turn to central extensions of 2-groups; we first recall some basic definitions from \cite{schommer2011central}. An \emph{extension} of a smooth 2-group $G$ by another smooth 2-group $A$ is the following set of data:
\begin{itemize}
    \item A smooth 2-group $H$,
    \item homomorphisms $f:A\rightarrow H$ and $g:H\rightarrow G$,
    \item A 2-homomorphism $\phi:gf\rightarrow 0$,
\end{itemize}
such that the $H$, endowed with a smooth $A$-stack structure via $f$, is an \emph{$A$-principal bundle}, i.e. locally trivial as a smooth $A$-stack. When $A$ is abelian, any extension $H$ of $G$ by $A$ induces a homomorphism $\phi_H:G\rightarrow \mathrm{Aut}(A)$ uniquely up to 2-homomorphism, where $\mathrm{Aut}(A)$ denotes the automorphism 2-group of $A$. We say that the given extension is \emph{central} if $\phi_H$ is isomorphic to the trivial homomorphism.

Then it follows from \cite[Theorem 99]{schommer2011central} that isomorphism classes of central extensions of a Lie group $G$ by $BA$, where $A$ is an abelian Lie group and $BA$ is the Lie 2-group $A\rightrightarrows pt$, are completely classified by elements of the differential group cohomology $H^3_{diff}(G,A)$. When $G$ is a compact Lie group and $A=U(1)$, then we have from \cite[Proposition 1.5]{brylinski2000differentiable} that
\[
H^n _{diff}(G,U(1)) \simeq H^{n+1}(BG;\mathbb{Z}).
\]
It then follows that, if $G$ is a simply-connected simple compact Lie group, then we have
\[
H^4_{sing}(BG;\mathbb{Z})\simeq \pi_3(G)\simeq \mathbb{Z},
\]
so central extensions of $G$ by $BU(1)$ are classified by integers. 

We will also consider the case when $G=U(1)$. In this case, we still have
\[
H^4(BG;\mathbb{Z})\simeq H^4(\mathbb{C}P^\infty;\mathbb{Z})\simeq \mathbb{Z},
\]
so central extensions of $U(1)$ by $BU(1)$ are still classified by integers. However, since $U(1)$ is not simply-connected, we need a different construction (which is much easier than the simply-connected case).

\subsection{Smooth string 2-groups as Lie 2-groups}

A big advantage of dealing with smooth 2-groups is that it is easy to find explicit finite-dimensional models of central extensions of Lie groups $G$ by $BA$ for abelian Lie groups $A$. Given a cocycle $\lambda=(\lambda^{3,0},\lambda^{2,1},\lambda^{1,2},\lambda^{0,3}) \in H^3_{diff}(G,A)$, one can explicitly describe the corresponding central extension in terms of  given data, i.e. $G$, $A$, and $\lambda^{i,3-i}$.

However, in order to deal with explicit descriptions of 2-connections on principal 2-bundles, it is much easier to use the notion of Lie 2-groups as we have to perform a Lie differentiation to get its Lie 2-algebra. Although the definitions of Lie 2-groups and smooth 2-groups are quite different to each other, in our cases, it is possible to transform smooth 2-groups which arise from central extensions of simply-connected compact Lie groups $G$ by $BU(1)$ to explicitly described Lie 2-groups. The details of this are not needed for our purposes as the resulting Lie 2-algebra is already known and takes a very simple form of a 2-term $L_\infty$-algebra. Hence we refer the readers to \cite{demessie2017higher} for details.

\section{Green--Schwarz mechanism and 2-connections for central $BU(1)$-extensions}
\label{sec:GS}

\subsection{Smooth 2-groups and 2-connections}
Gauge theory, in physics and also in mathematics, is not just about Lie groups, but also about principal bundles and connections. The concepts of principal $G$-bundles and $G$-connections, where $G$ is a Lie group, can be generalized to the case when $G$ is a smooth 2-group; in fact, a complete framework in the most general case when $G$ is a smooth $\infty$-group is also readily available; see \cite{nikolaus2015principal,nikolaus2015principal2} for more details. However, since introducing their definitions is not really necessary, we will omit the definition of principal 2-bundles and only work in the local case, i.e. trivial bundles. Since recent physical literatures regarding smooth 2-groups are about infinitesimal 2-group symmetries on QFTs, which translates to actions of infinitesimal gauge transformations on certain 2-connections obtained via physical arguments \cite{cordova2019exploring}, we will briefly recall the local descriptions of 2-connections and their transformations under infinitesimal 2-group gauge transformations.

In the classical case when $G$ is a Lie group, working locally and thus making the given principal bundle trivial allows us to define principal $G$-connections as $\mathfrak{g}$-valued 1-forms on a local Euclidean patch, where $\mathfrak{g}$ denotes the differentiation of $G$, i.e. $\mathrm{Lie}(G)$. In this case, the infinitesimal gauge transformation acts on connection 1-forms by
\[
A\mapsto A+dg,\quad g:U\rightarrow \mathfrak{g}.
\]

The case of 2-connections can also be dealt in a similar way, using the Lie 2-group model. In particular, given a Lie 2-group $G$, its differentiation is given by the moduli space of functors from manifolds to descent data of principal $G$-bundles with respect to the surjective submersion $N\times \mathbb{R}^{0\vert 1}\rightarrow N$. The result we get is an $L_\infty$-algebra (up to equvalence).

Let $G$ be a smooth 2-group, whose Lie algebra is equivalent to a 2-term $L_\infty$-algebra $\mathfrak{g}=(\mathfrak{g}_0\xrightarrow{\mu_1} \mathfrak{g}_1)$ (together with $L_\infty$-operations $\mu_2$ and $\mu_3$). Then a 2-connection on the trivial principal $G$-bundle on a local Euclidean patch $U$ is described by a pair $(A,B)$, where $A$ is a $\mathfrak{g}_1$-valued 1-form on $U$ and $B$ is a $\mathfrak{g}_2$-valued 2-form on $U$. The infinitesimal gauge transformation is then described as 
\[
\begin{split}
A &\mapsto A+dg+\mu_2(A,g)+\mu_1(\Lambda), \\
B &\mapsto B-d\Lambda -\mu_2(A,\Lambda)+\mu_2(B,g)+\frac{1}{2}\mu_3(g,A,A),
\end{split}
\]
for $\mathfrak{g}_1$-valued 0-forms $g$ and $\mathfrak{g}_0$-valued 1-forms $\Lambda$. 

\subsection{2-connections for central $BU(1)$-extensions: the case of $SU(n)$}

We finally consider explicit examples, which are central $BU(1)$-extensions of Lie groups, namely $U(1)$ and $SU(n)$ for $n\ge 2$. Since $SU(n)$ is simple, compact, and simply-connected, we know that 
\[
H^4(BSU(n);\mathbb{Z})\simeq \mathbb{Z}.
\]
Also, since $BU(1)\simeq \mathbb{C}P^\infty$, we also have
\[
H^4(BU(1);\mathbb{Z})\simeq \mathbb{Z}.
\]
Hence it follows that central extensions of $G$ by $BU(1)$, where $G$ is either $U(1)$ or $SU(n)$ for $n\ge 2$, are classified by integers, which we denote by $\kappa$ since it corresponds to \emph{level} of $G$-Chern--Simons theory. However, since $SU(n)$ is simply-connected and $U(1)$ is not, the explicit constructions of their central extensions are very different. 

We deal with the case of $SU(n)$ first. Then it is known that the Lie algebra of the resulting central extension, which we denote by $G_{n,\kappa}$, have a differentiation equivalent to the following Lie 2-algebra $\mathfrak{g}_{n,\kappa}$:
\[
\mathfrak{g}_{n,\kappa}=(\mathfrak{u}(1)\xrightarrow{\mu_1 = 0} \mathfrak{su}(n)).
\]
Here, the nonzero parts of $\mu_2$ and $\mu_3$ are given as
\[
\begin{split}
\mu_2(x_1,x_2) &= [x_1,x_2],\quad x_1,x_2\in \mathfrak{su}(n), \\
\mu_3(x_1,x_2,x_3) &= \kappa\left< x_1, [x_2,x_3] \right> \quad x_1,x_2,x_3\in\mathfrak{su}(n),
\end{split}
\]
where $\left< -,-\right>$ denotes the Killing form on $\mathfrak{su}(n)$. Hence a 2-connection for a trivial principal $G_{n,\kappa}$-bundle over a local Euclidean patch $U$ can be described as a pair $(A,B)$, where $A$ is an $\mathfrak{su}(n)$-valued 1-form on $U$ and $B$ is an $\mathfrak{u}(1)$-valued 2-form on $U$. Infinitesimal gauge transformations are given by 
\[
\begin{split}
    A &\mapsto A+dg, \\
    B &\mapsto B-d\Lambda + \frac{\kappa}{8\pi} \mathrm{Tr}( g [A,A] ).
\end{split}
\]

This is very close to \cite[Equation 1.24]{cordova2019exploring}, obtained via Green--Schwarz mechanism, except that after replacing $\Lambda$ by $-\Lambda$, we have $-\frac{1}{2}[A,A]$ in place of $dA$. Since we have
\[
F_A = dA + \frac{1}{2}[A,A],
\]
we see that \cite[Equation 1.24]{cordova2019exploring} defines a 2-connection exactly when $A$ is a flat $SU(n)$-connection.

The flatness condition of $A$ is directly connected to the notion of parallel 2-transports for 2-connections. Given a 2-connection $(A,B)$, its \emph{curvature} consists of a 2-form component $F_2$, which we call as the \emph{fake curvature}, and a 3-form component $F_3$, defined as
\[
\begin{split}
    F_2 &= dA + \frac{1}{2}\mu_2(A,A) - \mu_1(B), \\
    F_3 &= dB + \mu_2(A,B) - \frac{1}{6}\mu_3(A,A,A).
\end{split}
\]
In the case of the Lie 2-algebra $\mathfrak{g}_{n,\kappa}$, this becomes:
\[
\begin{split}
    F_2 &= dA + \frac{1}{2}[A,A] = \text{curvature form of }A, \\
    F_3 &= dB - \frac{1}{6}\mathrm{Tr}{A\wedge A\wedge A}.
\end{split}
\]
When the fake curvature $F_2$ (which is same as the curvature form of $A$) vanishes, then the given 2-connection $(A,B)$ defines a well-defined notion of a 2-monodromy along surfaces, just as how monodromies along curves are defined for (1-)connections. It is interesting to note that, under the condition that $A$ is flat, the 3-form curvature $F_3$ is $dB$ minus a scalar multiple of the Chern--Simons form of $A$.

\subsection{The case of central $BU(1)$-extensions of $U(1)$}
Now we discuss the case of remaining case, i.e. 2-connections when the gauge group is a central $BU(1)$-extension of $U(1)$. We denote the resulting central extension as $G_{1,\kappa}$, which corresponds to the cohomology class in $H^4(BU(1);\mathbb{Z}) \simeq \mathbb{Z}$ given by the integer $\kappa$. In this case the model we used in the case of $SU(n)$ does not work since $U(1)$ is not simply-connected, and thus we have to use a different model for the Lie 2-algebra $\mathfrak{g}_{1,\kappa}$ of $G_{1,\kappa}$. Fortunately, unlike the $SU(n)$ case, it is possible to describe $G_{1,\kappa}$ explicitly as a smooth 2-group induced by a crossed module. We can then simply differentiate it to compute $\mathfrak{g}_{1,\kappa}$.

A detailed description of $G_{1,\kappa}$ involving several different models of smooth 2-groups is given in \cite{ganter2018categorical}; we outline one of its descriptions here for the sake of self-containedness. Recall that a crossed module consists of two Lie groups $G,H$, together with an action of $G$ on $H$ by automorphisms and a smooth homomorphism $H\rightarrow G$, satisfying several properties. We consider the Lie groups $\mathbb{R}$ and $\mathbb{Z}\times U(1)$, where $\mathbb{R}$ acts on $\mathbb{Z}\times U(1)$ by 
\[
x\cdot (m,z) = (m,z\cdot \exp(\kappa mx))
\]
and we have a group homomorphism $\mathbb{Z}\times U(1)\rightarrow \mathbb{R}$ given by 
\[
(m,z)\mapsto \kappa m.
\]
Thus the Lie 2-algebra $\mathfrak{g}_{1,\kappa}$ is invariant of $\kappa$; it is given by 
\[
\mathfrak{g}_{1,\kappa} \simeq (\mathfrak{u}(1)\rightarrow \mathbb{R})
\]
where the $L_\infty$ operations $\mu_1,\mu_2,\mu_3$ vanish altogether. In other words, we have
\[
\mathfrak{g}_{1,\kappa}\simeq \mathrm{Lie}(U(1)\times BU(1)) = \mathfrak{u}(1)\oplus \mathfrak{bu}(1).
\]
It follows that infinitesimal gauge transformations for 2-connections on principal $G_{1,\kappa}$-bundles does not depend on the integer $\kappa$. In particular, for a 2-connection $(A,B)$, infinitesimal gauge transformations are described as
\[
\begin{split}
    A &\mapsto A+dg,\\
    B &\mapsto B-d\Lambda.
\end{split}
\]
As in the case of $SU(n)$, this formula matches the description of Cordova--Dumitrescu--Intriligator \cite[Equation 1.14]{cordova2019exploring} precisely when $A$ is flat. However, the difference between the $SU(n)$ case is that the flatness assumption on $A$ simply eliminates all terms involving the integer $\kappa$. Since the contribution of $\kappa$, obtained via Green--Schwarz mechanism, is solely contained in the term $\kappa F_A$, the fact that the formula gives the 2-connection only when $F_A=0$ suggests that Green--Schwarz mechanism might not be directly related to gauge theories when the gauge group is a smooth 2-group.

\section*{Acknowledgements}
The authors would like to thank Jacques~Distler and Yong-Geun~Oh for discussions and Craig~Lawrie and Jaewon~Song for much encouragement on writing this paper. The authors would like to thank the Center for Geometry and Physics at the Institute of Basic Sciences for the hospitality during the earlier stage of this project. M.~J.~K.~is supported by the U.~S.~Department of Energy, Office of Science, Office of High Energy Physics, under Award Numbers DE-SC0011632 and DE-SC0013528, and QuantISED Award DE-SC0020360, and a Sherman Fairchild Postdoctoral Fellowship.  S.~K.~is supported by Titchmarsh Fellowship.

\appendix
\section{$L_\infty$-algebras}

Recall that, given a Lie group $G$, its tangent space at the identity $1\in G$ gives a vector space, and differentiating adjoint representation $\mathrm{Ad}:G\rightarrow \mathbf{Aut}(G)$ gives a skew-symmetric bilinear operation on the vector space. This structure gives the Lie algebra $\mathrm{Lie}(G)$ of $G$. Since we are dealing with smooth 2-groups, it is natural to ask whether this differentiation construction is possible for smooth 2-groups and which structure are they naturally endowed with. The answer is given by Lie 2-algebras; in this section, we will discuss them, as well as its most general version, $L_\infty$-algebra.

We say that a graded vector space $\mathfrak{g}=\mathfrak{g}_0 \oplus \mathfrak{g}_1 \oplus \cdots$ is an \emph{$L_\infty$-algebra} if it is endowed with degree $k-2$ linear operations (called \emph{$L_\infty$-operations})
\[
\mu_k:\mathfrak{g}^{\otimes k}\rightarrow \mathfrak{g},
\]
sometimes also denoted $[-,\cdots,-]_k$, such that the following identity, called \emph{strong homotopy Jacobi identity}, holds for any elements $v_i \in \mathfrak{g}_{\vert v_i \vert}$:
\[
\sum_{i+j=n+1} \sum_{\sigma\in\mathrm{Shuffle}(i,j-1)} \chi(\sigma,v_1,\cdots,v_n) (-1)^{i(j-1)} \mu_j(\mu_i(v_{\sigma^{-1}(1)},\cdots,v_{\sigma^{-1}(i)}),v_{\sigma^{-1}(i+1)},\cdots,v_\sigma^{-1}(n)).
\]
Here, $\chi(\sigma,v_1,\cdots,v_n) \in \{1,-1\}$ is the $(v_1,\cdots,v_n)$-graded signature of $\sigma$ and $\mathrm{Shuffle}(p,q)$ denotes the set of $(p,q)$-shuffles, i.e. permutations of $\{1,\cdots,p+q\}$ such that
\[
\sigma(1)<\cdots<\sigma(p)\text{ and } \sigma(p+1)<\cdots<\sigma(p+q).
\]
This structure naturally occurs when one tries to differentiate (i.e. take the 1-jet) a smooth $\infty$-group using the Kan simplicial manifold model. For a more detailed discussion of this approach, see \cite{severa2006l_infinity}.

Things become much more explicit when we deal with $L_\infty$-algebras which arise from weak Lie 2-groups. When $\mathfrak{g}$ is a two-term $L_\infty$-algebra, we say that it is a \emph{Lie 2-algebra}. In that case we have $\mathfrak{g}=\mathfrak{g}_0\oplus \mathfrak{g}_1$, where the $L_\infty$-operations take the form
\[
\begin{split}
\mu_1 &: \mathfrak{g}_1\rightarrow \mathfrak{g}_0, \\
\mu_2 &: \mathfrak{g}_0 \otimes \mathfrak{g}_0 \rightarrow \mathfrak{g}_0, \\
\mu_2 &: \mathfrak{g}_1 \otimes \mathfrak{g}_0 \rightarrow \mathfrak{g}_1, \\
\mu_3 &: \mathfrak{g}_0 \otimes \mathfrak{g}_0 \otimes \mathfrak{g}_0 \rightarrow \mathfrak{g}_1.
\end{split}
\]
These operations should satisfy the following conditions:
\begin{itemize}
    \item $\mu_2$ on $\mathfrak{g}_0\otimes \mathfrak{g}_1$ is $\mu_1$-linear, i.e. $$\mu_2(x,\mu_1(y)) = \mu_1(\mu_2(x,y));$$
    \item $\mu_2$ on $\mathfrak{g}_0\otimes \mathfrak{g}_0$ satisfies the Jacobi identity up to homotopy given by $\mu_3$, i.e. $$\mu_2(x,\mu_2(y,z))+\mu_2(y,\mu_2(z,x))+\mu_2(z,\mu_2(x,y))=\mu_1(\mu_3(x,y,z));$$
    \item We have 
    \[
    \begin{split}
    & \mu_2(\mu_3(w,x,y),z)+\mu_2(\mu_3(w,y,z),x)+\mu_3(\mu_2(w,y),x,z)+\mu_3(\mu_2(x,z),w,y) \\
    &= \mu_2(\mu_3(w,x,z),y)+\mu_2(\mu_3(x,y,z),w)+\mu_3(\mu_2(w,x),y,z) \\
    &+ \mu_3(\mu_2(w,z),x,y)+\mu_3(\mu_2(x,y),w,z)+\mu_3(\mu_2(y,z),w,x).
    \end{split}
    \]
\end{itemize}
It is easy to see that this definition is directly related to the definition of weak Lie 2-groups. In particular, the last condition comes from the pentagon axiom. For a more detailed discussions, see \cite{baez2003higher}.

Note that the definition of Lie 2-algebras has nothing which relates to the unitor map in the definition of weak Lie 2-groups. Thus it is natural to say that Lie 2-algebras are first-order infinitesimal models to \emph{semistrict} weak Lie 2-groups, i.e. the ones having trivial unitors. However, although the weak Lie 2-group model for nontrivial central $BU(1)$-extensions of $SU(n)$ (or, in general, compact simply-connected simple Lie groups) are not semistrict, they still have semistrict Lie 2-algebras, which makes our discussions much easier. Indeed, given a simple compact Lie algebra $\mathfrak{g}$ and its Lie integration $G$, the Lie 2-algebras for nontrivial central $BU(1)$-extensions of $G$ are given by 
\[
\mathfrak{string}(\mathfrak{g})=\mathfrak{u}(1)\oplus \mathfrak{g},
\]
where $\mu_1=0$, $\mu_2$ is zero on $\mathfrak{u}(1)\otimes \mathfrak{g}$ and the Lie bracket of $\mathfrak{g}$ on $\mathfrak{g}\otimes \mathfrak{g}$, and $\mu_3$ is given by
\[
\mu_3(x,y,z)=\left< [x,y],z \right>,
\]
where $\left< -,- \right>$ denotes a Killing form on $\mathfrak{g}$.

\bibliographystyle{sortedbutpretty}
\bibliography{ref}
\end{document}